\documentclass[aps,prl, twocolumn]{revtex4}
\usepackage{amsmath}
\usepackage{epsfig}
\usepackage{amssymb}
\usepackage{dcolumn}
\usepackage{graphicx}
\usepackage{dcolumn}

\usepackage{color}

\newcommand{\beq}{\begin{equation}}
\newcommand{\eeq}{\end{equation}}
\newcommand{\dpsi}{\delta \psi}

\newcommand{\diff}{\mathrm d}

\newcommand{\rv}{\mathbf{r}}
\newcommand{\vv}{\mathbf{v}}
\newcommand{\kv}{\mathbf{k}}

\newcommand{\F}{\mathbf{F}}

\newcommand{\dr}{{\rm d} \rv}
\newcommand{\dk}{{\rm d} \kv}

\begin{document}
\date{\today}

\title{Beyond superfluidity in driven non-equilibrium Bose-Einstein condensates}

\author{Florian Pinsker}

\affiliation{Clarendon Laboratory, University of Oxford, Parks Road, Oxford OX1 3PU, United Kingdom}
\email{florian.pinsker@physics.ox.ac.uk
}

\begin{abstract}
The phenomenon of superfluidity in open Bose-Einstein condensates (BEC) is analysed numerically and analytically. It is found that  a superfluid phase is feasible even above the speed of sound, when forces due to inhomogeneous non-equilibrium processes oppose the contributions of homogeneous processes. Furthermore a regime of accelerating impurities can be observed for particular pumping/decay strategies. All findings are derived within the complex Gross-Pitaevskii (GP) theory with creation and annihilation terms. Utilising this framework the effective force acting on an impurity as it moves with velocity $v$ through the open condensate can be calculated. The result shows that the force is continuously increasing/decreasing with increasing velocity starting from the state of zero motion at $v =0$, a property that can be traced down to the additional homogeneous annihilation/creation term in the extended GP model. Our findings stand in stark contrast to the concept of a topological phase transition to frictionless flow below a critical velocity as observed for equilibrium Bose-Einstein condensates analytically \cite{pita, Pinsk}, 
numerically \cite{Win} and for trapped atoms  experimentally \cite{gr, 2D1}. 

\end{abstract}

\maketitle

\section{Introduction}  What defines superfluidity of a many-body system?  The answer can be given in terms of a statement based on Landau's theory of superfluidity \cite{puta, Land, teo}:  Below a certain critical velocity, due to non-existing energetically affordable elementary excitations within the many-body quantum system, an impurity moves dissipationless through the superfluid state of matter. Here the concept of drag force acting on the impurity due to interactions of the impurity with the fluid as it excites the quantum system turns out to be key as measure of dissipation. 

Particularly an impurity which is moving with velocity $v$ through a fluid in its quantum mechanical ground state can cause transitions from the fluid's  ground state to excited states lying on the line $\varepsilon = pv$ in the energy-momentum space \cite{blui, teo}. However if the whole energy spectrum of the fluid is above this line, the motion of the impurity cannot excite the system. This implies the superfluid phase where the impurity moves without resistance through the ensemble of unexcited matter particles. Even when the line $\varepsilon = pv$ intersects the energy spectrum of the fluid in its ground state, transition probabilities to these states can be strongly suppressed due to Boson interactions or due to the nature of the external perturbing potential \cite{blui}. For all scenarios the drag force experienced by the impurity gives us a quantitative measure of the state of the fluid and below a critical velocity, if the fluid cannot be excited, the impurity experiences no drag \cite{pita, Pinsk} a phenomenon already envisaged in  the classic papers in the beginning of the $20$th century \cite{uno, due, tre, quattro, Lon}. 

Experimentally superfluidity has been unambiguously observed for the condensed state in various weakly interacting and dilute effectively Bose gases of atoms or molecules at ultra-low temperatures in the nano Kelvin range \cite{gr, evid, ketter, heat, 2D1} and even in strongly interacting gases of $^6$Li fermions \cite{gr}.  On the other hand, theoretically, the ideal Bose gas, i.e. a quantum gas without interactions at all states of motion obeys dissipation of energy. Only for certain cases of {\it interacting quantum systems} such as in the BE condensed phase with its {\it intrinsic nonlinearity} due to particle interactions between Bosons, we observe a phase of superfluidity. Here excitations are suppressed when an obstacle moves through the BE condensed phase  by the nonlinear self-interactions which results in absence of drag \cite{Win, frisch, pom, win}. More recently it has become clear that a mean-field analysis leaves out subtle quantum fluctuations \cite{draggi, do1,do2,do3}, which, if taken into account e.g. in a linear response framework \cite{blui} or by directly considering the quantum correction in a quantum Bogoliubov analysis \cite{do3}, give rise to non-classical drag forces even below the critical velocity due to mean field theory. However a rigorous analysis shows that the superfluid phase persists even on a quantum level \cite{Lych1, Lych2}.

Superfluidity can be tested experimentally and proven by considering the dissipationless flow around impurities \cite{win, Win, frisch, pom, diss}, even in the presence of quantum fluctuations \cite{draggi, Lych1,Lych2}. This has been done theoretically in the semiclassical GP framework by the means of a Bogoliubov analysis for point-like \cite{pita} and subsequently Gaussian \cite{Pinsk} weakly-interacting impurities. This analysis confirmed the existence of  a superfluid phase in the leading order contribution - the drag force vanishes below the non-zero critical velocity that is equal the speed of sound for weakly-interacting obstacles, which hold as well quantum mechanically \cite{Lych1}, while geometric features significantly alter its magnitude \cite{Pinsk}. The analytical insights on nonequilibrium systems presented here build on this type of analysis and will be supported by numerical integration.

Once a superfluid is put in motion other aspects associated with this state of matter emerge. For nonequilibrium quasi-Bose-Einstein condensates such as Polariton condensates in their lowest energy state \cite{kasp} the low scattering rates from defects moving at velocities below the speed of sound and  the generation of Cherenkov-type waves at supersonic velocities have been observed experimentally \cite{exi}, while reduced drag at subsonic speeds has been noted in \cite{WouCa, Tim}. This reduced drag has been explained by the finite lifetime of Bogoliubov modes due to drain present in this open system \cite{WouCa}. In addition it has been realised that elementary excitations known from equilibrium condensates exist too - dark solitons are feasible in $1$d \cite{Hugo, da} or quantised vortices in $2$d non-equilibrium  condensates \cite{vor, vort} above a specific critical velocity or even spontaneously due to purely non-equilibrium dynamics \cite{Baer}.  The mathematical extension of the governing partial differential equation for open BEC in those scenarios implies intrinsic adaptions of the excitations' mathematical form \cite{Hugo, Pinsker11, Hugo} (see e.g. \cite{F,F1,F2,F4,F5,F6, F7} for rigorous results on equilibrium condensates). However several findings suggest similar response to motion or obstacles in relative motion to the open system \cite{Hugo, vor, vort, Light}. Now in the semiclassical mean-field regime many Bosons in the ground state are described by solutions to the Gross-Pitaevskii equation (GPE)  \cite{Pita1, Gross, puta}, while  non-equilibrium condensates in their simplest form are described by an extended GPE with additional complex terms, which correspond to creation and annihilation operators of these modes \cite{Light, Wout, AtomLaser, exi}.

One goal of this paper is to point out implications of the non-equilibriumness of open systems on the possibility of leading order superfluidity by testing a quantum state in relative motion to an obstacle. Here we use the complex GP framework as model of the coherent many-body system, corresponding to many particles being in the same quantum state, and we study the behaviour of Dirac and finite-sized obstacles. Particles can enter and leave this macroscopically occupied coherent quantum state to which we refer to as the condensate wave function. So the number of particles in this macroscopic state is not preserved and the actual condensate wave function depends on the scattering of particles into this state and the particles' decay. For example if no particles are added to the condensate, the occupation number of particles in this mode eventually goes to zero. On the other hand, due to pumping particles into the condensate patterns might emerge \cite{Baer} and potentially new properties. While by naively applying the analogy to conserved BE condensed systems a superfluid phase aka absence of drag would be expected for small velocities below the speed of sound of the fluid due to particle interactions, while above a critical velocity a drag force would arise due to the possibility of emission of elementary excitations \cite{pita, Pinsk, exi, WouCa}. The continuous flow of particles into the condensate phase and the corresponding balancing drain may alter the scenario as studied numerically in \cite{WouCa} and thus superfluidity could be suppressed \cite{Sky}. On the other hand could we scatter particles into the condensate, such that the condensate wave function implies an effectively vanishing drag force aka superfluidity?

\subsection{Physical scenario}

So far the arrangement of pumping and the presence of decay has been considered in terms of analog behaviour to equilibrium BEC, e.g. the absence or reduction of scattering from an obstacle as it moves relativ to the condensate \cite{kaka}.  Here we particularly will consider an inserted obstacle to be co-moving with a particular pump distribution or equivalently both the pump and the obstacle are stationary and the surrounding Polariton quantum-fluid is passing by. We will study the effect of arranging the pumping distribution properly without having additional momentum, when both are moving with the same relative speed to the condensate. This will allow us to observe in what way the obstacle can be affected by the non-equilibrium pumping distribution.

To elucidate those and superfluid aspects of an open interacting and coherent many-body system, let us introduce first the explicit theoretical framework we use as model of the condensate and subsequently we will derive by deduction key statements.

\section{Condensate wave equation}  The free energy of a BEC, which experiences the external potential $V$ and which has a self-interaction strength $g$, is given by \cite{Inge, puta, Pita1, Gross, Light, leg}
\begin{equation}\label{min}
\mathcal E [ \phi ] =   \int \omega (\kv) |\hat \phi (\kv)|^2 \dk +  \int \bigg(\frac{g}{2} |\phi|^4 +  V |\phi|^2 \bigg) \dr.
\end{equation}
Here we use the convention $\mathcal F (f) \equiv \hat{f}(\kv) =
\int_{\mathbb R^d} f(\rv)\, e^{-i \kv \cdot \rv}\, \diff \rv$,
which denotes the Fourier transform of $f(\rv) = \frac{1}{(2\pi)^d}\int_{\mathbb R^d} \hat f(\kv)\,
e^{i \kv \cdot \rv}\, \diff \kv$.
The dispersion of the condensate is $\omega (|\kv|): \mathbb R^d \to \mathbb R$  and for simplicity can be assumed to be parabolic $\omega (|\kv|) \sim k^2$ (with $|\kv| = k$) corresponding to GP theory or free particles \cite{mk2}.
Naturally GP theory is considered in  $3$d, however, reduction to lower dimensions occurs e.g. for strong confinement along a spatial dimension \cite{Inge}. In $3$d $g = 4 \pi \hbar^2 a /m$ and $\max_\rv V (\rv)= 4 \pi \hbar^2 b /m \equiv V_0$ are respectively the particle-particle and particle-impurity coupling, where $a$ and $b$ are the corresponding scattering lengths and $m$ is the effective mass \cite{pita}. For the $2$d case we have  $g_{2 {\rm d}} = \sqrt{2 \pi} \hbar^2 a/ a_z$ and here $a_z = \sqrt{\hbar / m \omega_z}$ is the oscillator length with $\omega_z$ being the trapping strength \cite{Inge, puta}.  The condensate wave function $\psi (\rv, t)$ is the minimiser of \eqref{min}, i.e. the mode in which all particles of the dilute weakly-interacting closed Bose gas condense \cite{puta, Inge}. Mathematically  $\psi(\rv, t): \mathbb R^{d+1} \to \mathbb C$ and by performing a variation of the energy $\mathcal E [ \psi ]$ with respect to $\psi^*$ under the norm/particle preserving constraint $\| \psi \|^2_2 = 1$, we get the Euler-Lagrange equation for the minimiser. The so-called Gross-Pitaevskii equation is
\begin{equation}\label{model}
 i \partial_t \psi (\rv,t) = q \star \psi (\rv,t) + \left( g |\psi|^2 + V (\rv, t)  - \mu \right) \psi (\rv,t).
\end{equation} 
$\omega(\kv) $ is the Fourier transform of $q$, iff the transform exists and we write $\mathcal F^{-1} ( \omega (\kv) \mathcal F (f)) \equiv (q \star  f) (\rv,t)$. 
The chemical potential in Eq. \ref{model}, i.e. the energy needed to add another particle, is $\partial \mathcal  E [ \psi ]/\partial N = \mu$ \cite{Inge}.  To account for the open non-equilibrium dynamics of the condensate \eqref{model} one generally considers an extension of the form
\begin{multline}\label{natural1}
i \partial_t  \psi (\rv,t) = q \star  \psi (\rv,t) +   i P  (\psi,\rv,t)  - i \Gamma_d  \psi (\rv,t) \\ + \bigg( g  |\psi|^2  + V (\rv, t) - \mu \bigg) \psi (\rv,t),
\end{multline}
which is applicable for atom laser systems \cite{AtomLaser} as well as for non-equilibrium Polariton condensates \cite{Light, exi}.
In \eqref{natural1} we consider additional physical parameters:  $\Gamma_d $ is the homogeneous decay (or pump) rate of condensed particles, while $P$ approximates the inhomogeneous non-equilibrium processes acting on the condensate fraction \cite{AtomLaser, Light}, which shall have an integrable Fourier transform. We assume that $P  (\psi,\rv,t) \simeq P' (\rv,t)  \psi (\rv,t)$, while we drop the superscript in what follows. Nonlinear density dependent processes can be regarded as adding a constant to the pump $P$, when considering a first order linear waves analysis. Examples of pump terms of this form are  e.g. spatially dependent incoherent scattering of reservoir particles into the condensate phase \cite{Light}. These terms have been used to describe experimental results of Polariton condensates in the mean-field regime \cite{alex, PRL,PRX} and see \cite{Baer, Light} for theoretical works on that matter.
 Due to the generalisation of the guiding equation to include non-equilibrium processes the norm of the wave function $\| \psi \|^2_2 = f(t)$ now varies in time. 

\section{Impurity waves} The impurity moving with velocity $\vv$ in the stationary fluid frame is modelled by the external potential $V (\rv, t)  =V_0 e^{- \frac{1}{2 \sigma^2} (\rv - \vv t)^2}$ \cite{Pinsk}, which has an amplitude $V_0$ and a width $\sigma$, or simply by a Dirac delta function $V_{\rm Dirac} (\rv, t)  =V_0 \delta (\rv - \vv t)$ \cite{pita}. These potentials model an inserted atom or a laser beam \cite{puta, Light, Pavloff}. We suppose the linear waves generated by these potentials are of the form $\psi = \phi_0 + \delta \psi$ \cite{pita, Pinsk, glad}, where $\phi_0$ represents the unperturbed part solving the complex GPE \eqref{natural1} without potential and $\dpsi (\rv ,t )$ denotes a small perturbation due to the presence of an impurity. By inserting this Ansatz in $\eqref{natural1}$ and dropping terms of order $\dpsi^2$ we get the Bogoliubov equation for this perturbation,
\begin{multline}\label{lin}
 i \frac{\partial \dpsi}{\partial t} = q \star  \dpsi + V \phi_0 + \\ + g \left(2 |\phi_0|^2 - \frac{\mu}{g} \right) \dpsi + g \phi_0^2 \dpsi^* + i  P  \phi_0  - i \Gamma_d  \dpsi.
\end{multline}
Here we restrict our consideration to small weakly-interacting impurities, $V \simeq \delta \psi$. Furthermore we utilise the identity $ \partial \dpsi(\rv - \vv t)/ \partial t = - \vv \nabla \dpsi(\rv - \vv t)$ \cite{pita, Pinsk, glad} and switch in the frame moving with the impurity, $\dpsi (\rv,t) = \dpsi (\rv - \vv t) = \dpsi (\rv')$, return to the notations without superscripts and in addition consider Eq. \ref{lin} in $\kv$-space.  Thus the wavefunction $\dpsi_\kv =  \int e^{-i \kv \rv } \dpsi \dr$ satisfies the Bogoliubov equation in $\kv$-space given by
\begin{multline}\label{lin2}
\kv \vv \dpsi_\kv = \omega (\kv) \dpsi_k + \int e^{-i \kv \rv } V \phi_0 \dr - i \Gamma_d  \dpsi_\kv \\ + i \int e^{-i \kv \rv } P \phi_0 \dr  +  g \left(2 |\phi_0|^2 - \frac{\mu}{g} \right) \dpsi_\kv + g \phi_0^2 \dpsi^*_{-\kv}.
\end{multline}
We note the phase factor identity, $\mu = g n$ with $n=  |\phi_0|^2$ and assume $\phi_0$ to be real valued for the sake of simplicity. Integrating the potential term e.g. $V (\rv, \sigma, V_0)$ in $2$d yields $2 \pi \sqrt{n} \sigma^2 V_0  e^{- \sigma^2 k^2}  \equiv f_{2 {\rm d} }(k^2)$ and so on and thus we get
 the algebraic equations of the form
\begin{multline}\label{lin5}
 \kv  \vv \dpsi_\kv = \omega(\kv) \dpsi_k + \mu \left( \dpsi_\kv + \dpsi^*_{-\kv} \right) + \\ + f_{2 {\rm d} }(k^2) + i  \mathcal F (\phi_0 P) - i \Gamma_d \dpsi_k.
\end{multline}
These are analytically solved,
\beq\label{solution}
\dpsi_\kv = - \frac{S_{2 {\rm d}}(i \Gamma_d + \kv \vv + \omega(\kv) ) + 2 i \mu {\mathcal Im} (S_{2 {\rm d}}) }{(\Gamma_d - i \kv \vv)^2 + 2 \mu  \omega (\kv) + \omega^2},
\vspace{1mm}
\eeq
where we make use of the notation $S_{2 {\rm d}} = f_{2 {\rm d} }(k^2) + i \mathcal F (\phi_0P)$.
Similar linear waves have been derived in several scenarios and we refer to \cite{pita, glad, Pinsk, mk2, Mar} for related results. Here the finite sized impurity is represented by a form factor $f_{2 {\rm d} }(k^2, \sigma, V_0)$ and the growth and decay terms alter the linear waves by extending the solution to the imaginary plane. The dimensional form factor is determined by the Fourier transform of the impurity and for general dimensions given by $f_{{\rm D}} = (2\pi)^{\frac{D}{2} }  \sqrt{n} \sigma^D V_0  e^{- \frac{D}{2}\sigma^2 k^2}$ \cite{Pinsk}. Moreover for the simple impurity $V_{\rm Delta}$, the form factor becomes $f_{{\rm Dirac}} = V_0 \sqrt{n}$. 

The energy spectra for the solutions \eqref{solution} are in general complex-valued and thus a naiv application of Landau's critical velocity based on real-valued dispersions \cite{Yukalov} cannot be applied, but we turn to the drag force as effective measure of friction.


\section{Drag Force} After deriving the form of the perturbed wave we investigate the force the impurity experiences as it moves through the non-equilibrium condensate.  The definition of the drag force is \cite{pita, draggi} 
\beq\label{defdrag}
\F = - \langle \Psi^\dagger |\nabla V(\rv,t)| \Psi  \rangle =  - \int |\psi|^2 \nabla V \dr,
\eeq
when assuming the mode of the Bose gas $\Psi^\dagger$ is fully described by the complex order parameter equation \eqref{natural1}. To calculate the drag force we employ the ansatz $\psi = \phi_0 + \dpsi$ and again neglect the $\dpsi^2$ terms. The result is
\begin{multline}\label{here2}
\F = 
 - 2(2\pi)^{\frac{D}{2} } \sigma^D V_0 \\ \int  \int \phi_0(\dpsi  + \dpsi^*) i \kv   e^{i \kv \rv - \frac{D}{2}  \sigma^2 k^2}  \frac{\dk}{(2\pi)^D} \dr.
\end{multline}
Above calculation includes the special case of the Dirac delta, when $V_0 \sim 1/ \sigma^D$ as $\sigma \to 0$, (see \cite{pita} for similar expressions).
The remainder in $\eqref{here2}$ is calculated as explicitly presented in the appendix. So after some complex algebra we obtain the effective semiclassical force acting upon the impurity in the non-equilibrium condensate,
\begin{widetext}
\begin{equation}\label{huhu}
\frac{\F }{c_{{\rm D}}} = \\  \int {\mathcal Re}\left(  e^{- \frac{D}{2}  \sigma^2 k^2}  \frac{S_{\rm D}(i \Gamma_d + \kv \vv + \omega(\kv) ) + i 2 \mu {\mathcal Im} (\mathcal F (i P \phi_0)) }{(\Gamma_d - i \kv \vv)^2 + 2 \mu  \omega (\kv) + \omega^2}  i \kv \right)    \frac{\dk}{(2\pi)^D},
\end{equation}
\end{widetext}
with $c_{{\rm D}}  = 2 \sigma^D (2 \pi)^{D/2} V_0 \sqrt{n} $. A similar result is obtained  as $V \to V_{\rm Dirac}$ by the substitution $c_{{\rm D}} \to 2 V_0 \sqrt{n}$ and $f_{{\rm D}}  \to f_{{\rm Dirac}}$. 

In all cases and independent of dimension, we observe that the denominator of the drag force in Eq. \ref{huhu} has a pole, if and only if
\beq
- (\Gamma_d - i \kv \vv)^2 \equiv 2 \mu  \omega(\kv)  +  \omega^2(\kv).
\eeq
This holds true for real-valued (free particle) dispersion relations $ \omega (\kv)$ only when $\Gamma_d = 0$, a case that corresponds to a BEC without homogeneous leakage, decay or constant gain of particles. The corresponding proofs of leading order existence of the superfluid phase transition for finite impurities and Delta impurities in equilibrium BEC was done in \cite{Pinsk}  and \cite{pita} respectively, where Sokhotsky's formula \cite{sok} was utilised to solve the integral with pole, thus implying the topological superfluid phase transition. The absence of a pole gives rise to a continuous (drag) force and correspondingly to the suppression of superfluidity when $\Gamma_d \neq 0$ as observed numerically in \cite{Sky} and analytically for a Dirac impurity in \cite{Mar}.  Physically particles leaving the condensed phase $\psi$ add drag to the superfluid phase.
 
\begin{figure}[ht]
\begin{tabular}{cc}
\begin{picture}(150,0)
\put(-12,-80) {\includegraphics[scale=0.37]{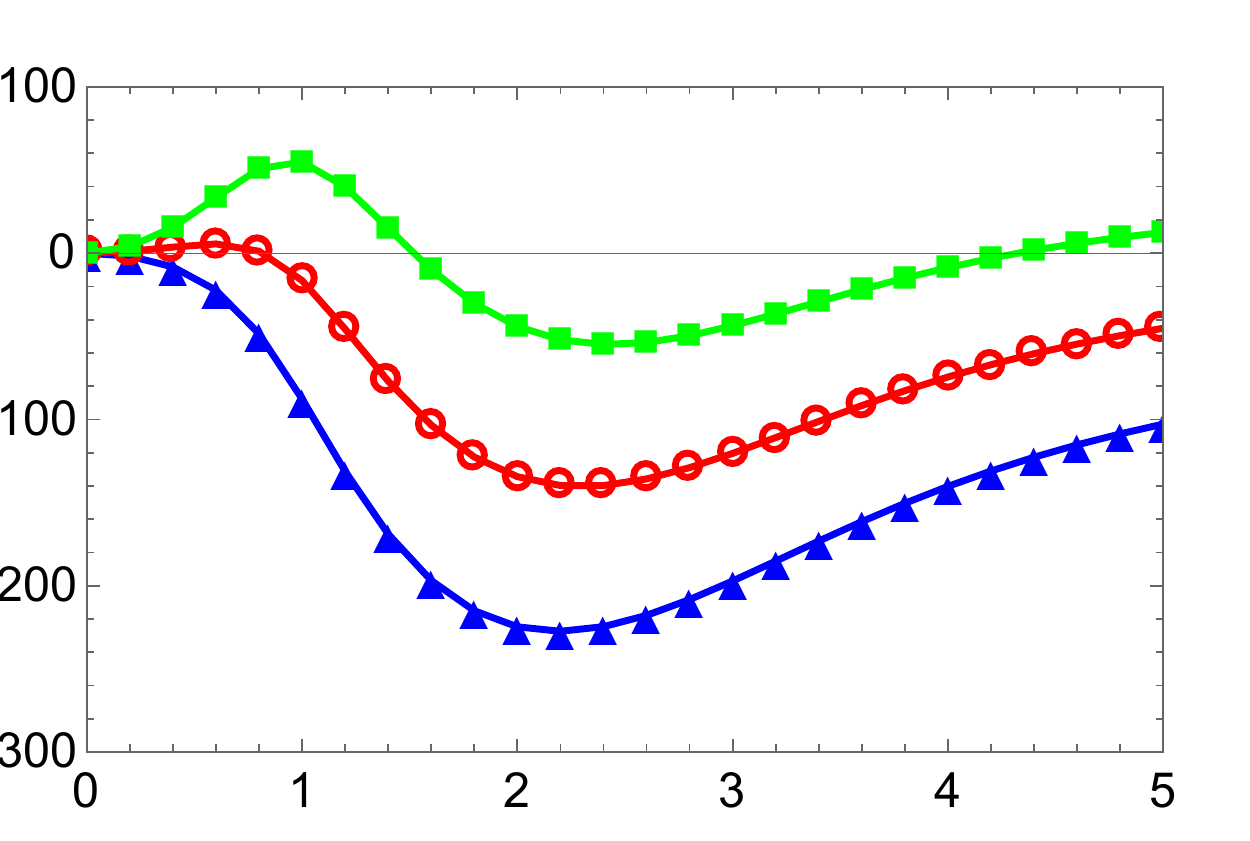} }
\put(90,-10){ \textcolor{black}{(a)}}
\put(-15,7){ \textcolor{black}{$F_v$}}
\put(50,-87){ \textcolor{black}{$v$}}
\end{picture} \hspace{2mm} &
\begin{picture}(155,-80)
\put(-44,-80) {\includegraphics[scale=0.37]{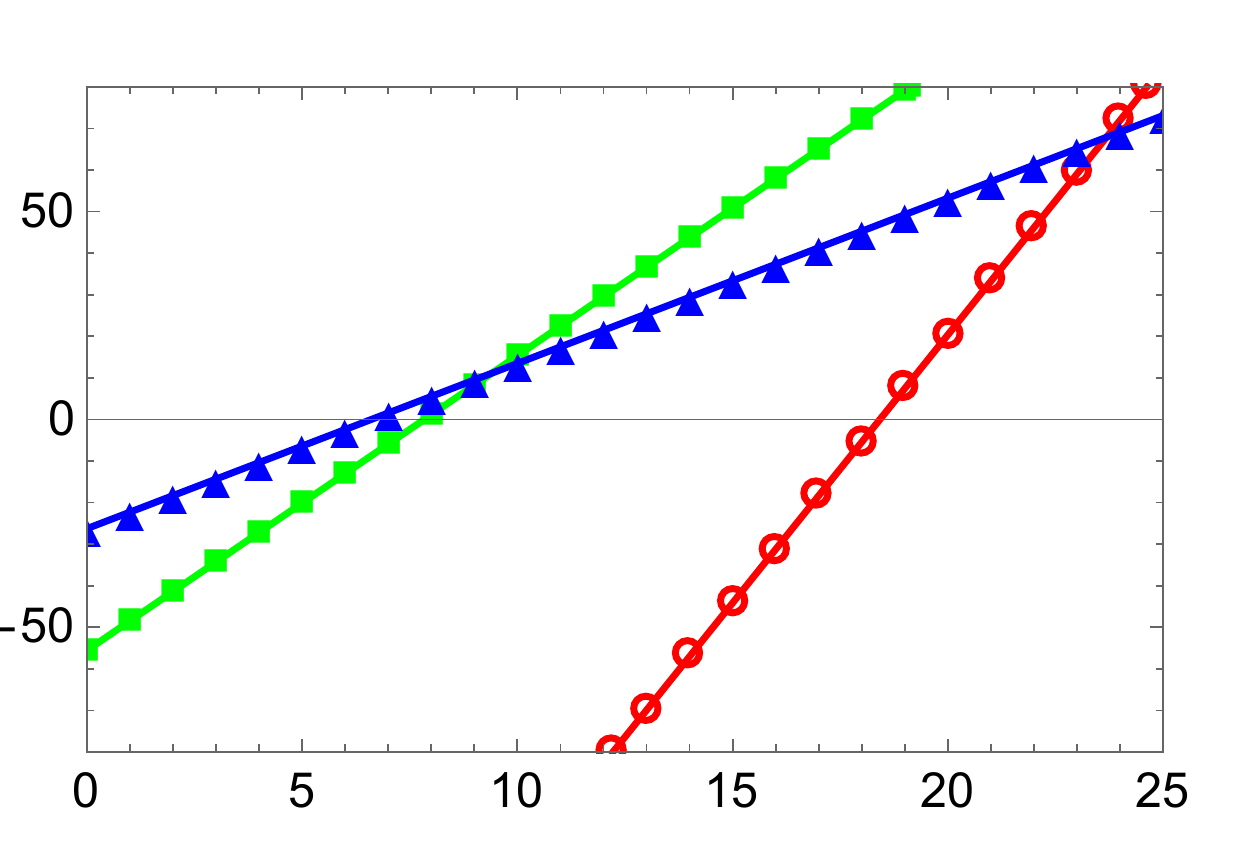} }
\put(63,-23){ \textcolor{black}{(b)}}
\put(-40,7){ \textcolor{black}{$F_v$}}
\put(15,-87){ \textcolor{black}{$P$}}
\end{picture}
\end{tabular}

\vspace{30mm}

\caption{ (a) Drag $F_v < 0$ and driving force $F_v > 0$ projected on $\vv$ for pumping strengths $P=1$ (triangles), $P=8$ (circles) and $P=15$ (squares).  Numerical parameters are given in \cite{Parameters}. In (b) the force $F_v $ is presented for the same numerical values however as a function of $P$ for fixed velocities,  $v=0.3$ (triangles), $v=0.4$ (squares) and $v=1$ (circles). The intersection of the graphs with the axis at $F_v =0$ can be identified with a superfluid regime.}
\label{F1}
\end{figure}
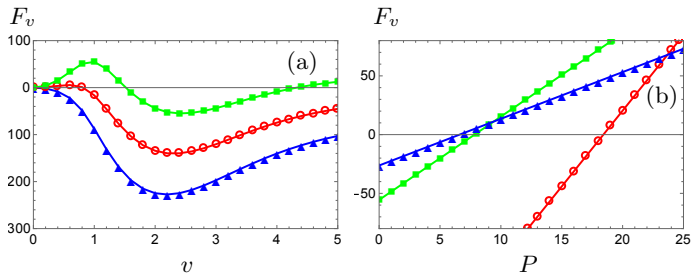

 On the other hand here we point out that for a vanishing numerator in \eqref{huhu} or integrals thereof we observe a superfluid phase and an acceleration regime when the effective sign of the force projected on the direction of motion is positive. In Fig. \ref{F1} (a) we numerically show the velocity dependence of the drag force acting on the Gaussian impurity in $2$d for various pumping strengths, given a pump of the form $P = P_0 \sqrt{n}  \delta (\rv - \vv t)$. While for low pumping strengths (triangles) the drag is persistent for all $\vv$ we note that the absolute drag is decreasing again (when the sign is still negative) for larger velocities as previously observed in equilibrium condensates for finite sized impurities \cite{Pinsk, Pavloff}. In contrast localised Dirac impurities yield a monotonically increasing amplitude of the drag force \cite{pita}. On the other hand for larger pumping strengths (circles and squares) we obtain forces with positive values, hence accelerating the impurity and when $F_v = 0$ at specific $v$ we effectively gain a superfluid phase due to balanced inhomogeneous and homogeneous non-equilibrium processes. Fig. \ref{F1} (b) shows the explicit (linear) dependence of the force on $P$ for fixed velocities and thus the critical threshold of superfluidity and the impurity accelerating regime.
 
Considering a semiclassical picture the local gain of particles into the condensed phase can cause the impurity to be pushed further by the pumping induced density variation, thus opposing/annihilating the drag created by potentially exciting the superfluid. In this sense scattering from and into the condensed phase (of particles even without a momentum as assumed by the mathematical form of the pumping term) act as additional external forces on the impurity due to their spatial distribution close to the impurity. Those add to the forces due to exciting the fluid and due to quantum fluctuations. A higher local density of the condensate at the impurity implies the possibility of larger scattering rates between the impurity and the condensate, which therefore provide the possibility of effectively accelerating the impurity through scattering. By considering the simple mathematical scenario of a $1$d Gaussian impurity potential and a step function density $|\psi|^2$ in $\eqref{defdrag}$, one can immediately conclude that adding density behind or in the wake of the impurity as it moves through the condensate will cause gain of momentum due to enhanced scattering probabilities in direction of motion. The scattered particles leave the condensate by gaining the momentum and thus induce via momentum conservation a gain of momentum of the impurity and effectively a force acting upon it. 

Furthermore note that even an infinitesimal small addition to the superfluid density implies an infinitesimal acceleration of the impurity. Interestingly we note that for a locally driven condensate without homogeneous leakage or pumping, i.e. $\Gamma_d = 0$, the impurity can be accelerated by this mechanism within a state of true superfluidity, i.e. below the speed of sound.

We point out that once the obstacle is accelerated by the pump, the pump has to be co-moving with the obstacle to maintain the acceleration of the obstacle. However in each stationary frame the pump does not induce additional momentum on the obstacle but is co-moving with the obstacle and merely modifies the density of the superfluid in the corresponding frame, which in turn accelerates the obstacle. In this sense the observed acceleration is a reversed drag force as discussed in \cite{pita}.

Now to calculate the continuous force acting upon an impurity analytically, when the various contributions in the numerator  \eqref{huhu}  do not cancel out for the explanatory case, $P=0$, we consider the projection of the drag force $\F$ on the velocity of the obstacle $\vv$, which we denote $F_v$.
By considering a parabolic kinetic dispersion $\omega$ and using the symmetry properties of the integrand, complex algebra and by approximating the integral by an expansion of the integrand to the quadratic order in $v$ we obtain a simple expression for the integral. In $2$d, which is the natural dimension e.g. for Polariton condensates, the result is a drag force acting in opposite direction of motion on the obstacle,
\begin{multline}\label{result}
\frac{F_v}{c_{2{\rm d}}} =  - \frac{\pi v^2 \Gamma_d}{2} \int^\infty_0  \frac{ e^{-2 \sigma^2 \rho} \rho^2 }{(\Gamma_d^2 + \rho^2 + 2 \rho \mu)^2}   \frac{d \rho }{(2\pi)^2} + {\mathcal O}(v^4).
\end{multline}
Even for small velocities of the impurity the drag force obeys approximately quadratic behaviour in $v$, i.e. a non-vanishing drag force, in particular since the integral in \eqref{result} is strictly positive. This contrasts the equilibrium BEC results \cite{pita, Pinsk}, where a critical velocity equal the speed of sound has been noted below which there is no drag at all corresponding to the superfluid phase. Logical consistency of the results presented here and those in \cite{pita, Pinsk} is given when $\Gamma_d \to 0$.

Next we solve the integral in \eqref{result} for the special case, $\Gamma_d \simeq \mu$, which yields the result
\begin{widetext}
\begin{equation}
\frac{F_v}{c_{2{\rm d}}} \simeq  - \frac{\pi v^2 \Gamma_d}{6 \mu}   \left(1 + t
    \mu' \left(5 + 2 \mu' + 2 e^{2 \mu'} (3 + 2  \mu' (3 + \mu')) \right) \left({\rm Chi}[2 \mu'] - \log[\sigma^2] - {\rm Shi}[2 \mu'] \right) \right),
\end{equation}
\end{widetext}
where we have used the abbreviation $\mu' \equiv \mu \sigma^2$ and the notation ${\rm Chi}[z] = \gamma + \log(z) + \int^z_0 (\cosh(t) -1)/t dt$, where $\gamma \simeq 0,5772$ is Euler's constant and  ${\rm Shi}[z] = \int^z_0 \sinh(t)/t dt$.
An additional example for a different parameter regime of Eq. \ref{result} is presented in the supplemental material, where again we confine the consideration to the quadratic order in velocity approximation. 

When neglecting the geometry of the impurity, i.e. $V \to V_{\rm Delta}$, and by considering the quadratic order in $v$ approximation, while again confining our consideration to $2$d, we directly obtain the formula for the drag force, 
\begin{multline}\label{supi}
 \frac{ F_v}{c_{\rm Dirac}}= - \frac{\pi v^2 \Gamma_d}{2} \int^\infty_0  \frac{  \rho^2 }{(\Gamma_d^2 + \rho^2 + 2 \rho \mu)^2}   \frac{d \rho }{(2\pi)^2} = \\ = - \frac{v^2 \Gamma_d}{32 \pi} \frac{\sqrt{d} \left(\frac{ (1+d) \pi}{\sqrt{d}} -2 \right) - 2 (1+d){\rm cot}^{-1} [\sqrt{d}]}{ d^{3/2}},
\end{multline}
given $d \equiv \Gamma_d^2  - \mu^2 > 0$. The drag force is decreasing with increasing $d$ approaching zero when $d \to \infty$, hence suggesting an extremal superfluid regime for this special case, while finite $d$ obey a quadratic in $v$ dependence of its magnitude and when $d = 0$ we obtain the finite drag force $F_v=   - c_{\rm Dirac} v^2/(48\pi)$. The linear Schr\"odinger equation in the scenario of a quantum fluid flowing past an impenetrable cylindrical obstacle of radius $R$, obeys a drag law for high velocity or large object size, i.e. $v\gg \hbar/mR$, that approaches the classical limit as well \cite{Win}, i.e. $F^{\rm ideal}_{v}\,= - c \rho_0 R v^2$, where $\rho_0$ is the density of the fluid and $c$ a dimension dependent constant. Although the results \eqref{result} and \eqref{supi} show analog approximatively quadratic behaviour, the drag force increases less than quadratic for $v\ll \hbar/mR$ for the ideal Bose gas  \cite{Win}. 


\section{Inhomogeneous pumping} Finally we turn to estimate the pumping terms in a $2$d scenario. We suppose $P = P_0 \sqrt{n}  \delta (\rv - \vv t)$ and thus when switching in the moving frame we have $\mathcal F (P) = P_0 \sqrt{n} \in \mathbb R$, i.e. a Dirac function pumping spot moving with the impurity, while we note that the complementary extreme case, i.e. a homogeneous pumping spot would simply adapt $\Gamma_d$ in the previous considerations. For the sake of conciseness we consider a Dirac delta impurity and obtain in the quadratic order of $v^2$ an inhomogeneous force
\begin{multline}\label{resultpump}
\frac{F^{\rm Inhom}_v}{c_{\rm Dirac}} =   \frac{ P_0 \sqrt{n}  \Gamma^2_d v^2 \pi}{8} \cdot \\ \left(\frac{2}{d} -\left(\frac{1}{d} \right)^{3/2} \mu \pi + \frac{2 \mu \tan^{-1} (\mu/\sqrt{d})}{d^{3/2}}  \right),
\end{multline}
using $d \equiv \Gamma_d^2  - \mu^2 \neq 0$ with more details presented in the appendix. We note that the inhomogeneous force can be directed along the direction of motion and thus is opposing the drag created by the homogeneous pumping/decay terms for specific $\mu$ and $\Gamma_d$, hence showing analytically in particular the possibility of a balance between homogeneous drag and local driving and thus a non-equilibrium superfluid phase which in the quadratic order order is independent of velocity. Furthermore increasing $P_0$ above this critical value implies an acceleration of the impurity that is proportional to $P_0$.


\section{Discussion} By utilising Bogoliubov's perturbation theory applied to the scenario of a (finite sized) Gaussian and a Delta function impurity moving through a non-equilibrium Bose-Einstein condensate we have obtained the formulas for the drag forces acting opposite the direction of motion on the impurities and the opposing driving forces. In general movement with velocities larger than the speed of sound leads to a non-zero drag force due to Cherenkov radiation of phonons as previously noted for the Dirac and Gaussian impurities in \cite{pita} and \cite{Pinsk} respectively.  In stark contrast, however, we have observed  that the drag force is non-vanishing as soon as the impurity is set in motion due to a constant non-equilibrium term in the governing equation, which confirms the numerics in \cite{Sky}.  Furthermore it has been shown that the force depends on the width and amplitude of the moving obstacle in the stated analytical form. We point out that the presented analysis does not include the effective drag due to nonlinear excitations such as vortices, vortex rings, solitary waves or solitons, which add energy dissipation and thus cause additional drag to the impurity. When nonlinear excitations are absent our mathematical analysis clarifies the linear waves contribution to the (drag) force acting on the weakly-interacting impurity as it moves at any velocity through the open BEC.

Now, if a balance between homogeneous and inhomogeneous non-equilibrium contributions is given a superfluid phase, i.e. a vanishing drag force in the stationary reference frame is feasible even above the speed of sound as shown numerically and analytically.  The most significant insight however is that the impurity can be accelerated by the inhomogeneous terms indicating a regime of the non-equilibrium condensate beyond that of superfluidity. The physical situation can be understood e.g. by a generalised Lagrangian formalism \cite{Pinsker11}, which treats the complex non-equilibrium terms as external forces acting on the condensate wave function. By direct inspection of \eqref{defdrag} one observes that for a Gaussian shaped impurity adding amplitude behind the moving impurity will imply an acceleration due to increased likelihood of scattering between the impurity and the condensate and the possible gain of momentum. The non-equilibrium terms, as presented here, can cause drag as well as an effective acceleration of the impurity when the pump is co-moving with the impurity. Finally using a more subtle approximation of quantum fluids by taking into account quantum fluctuations suggests that there may be additional dissipation \cite{draggi} even in closed BEC. However, those fluctuations can - on average - be balanced or even revresed by inhomogeneously driven forces as shown in this paper. 


\section{Acknowledgements} F.P. acknowledges financial support through his Schr\"odinger Fellowship (Austrian Science Fund (FWF): J3675) at the University of Oxford and the NQIT project (EP/M013243/1).

\newpage

\section{Appendix}

\subsection{Calculating the drag force without constant gain/loss}

To calculate the drag force we rely on the following calculation. For the sake of simplicity we just consider the case in $2$d here while presenting the formulas for the general dimensions in the main text. Using the definition of the drag force, Eq. $8$ in the main text, and by switching into momentum space we get,
\begin{multline}\label{here}
\F = -2 \sigma^2 \pi V_0 \int |\phi_0 + \dpsi|^2 \vec \nabla   \int e^{i \kv \rv} e^{-\frac{1}{2} \sigma^2 k^2}  \frac{\dk}{(2\pi)^2} \dr\\ \simeq -2 \sigma^2 \pi V_0 \int  \int (\phi_0^2 + \phi_0(\dpsi  + \dpsi^*)) i \kv   e^{i \kv \rv} e^{-\frac{1}{2} \sigma^2 k^2}   \frac{\dk}{(2\pi)^2} \dr  \\ = \F_0 \\ -2 \sigma^2 \pi V_0 \int  \int \phi_0(\dpsi  + \dpsi^*) i \kv   e^{i \kv \rv-\frac{1}{2} \sigma^2 k^2}  \frac{\dk \dr}{(2\pi)^2},
\end{multline}
by neglecting $\dpsi^2$ terms. Here we have defined $\F_0$ in the last step. To determine the remainder in $\eqref{here}$ for $\dpsi$, we use the following calculation:
\begin{widetext}
\begin{multline}\label{ho}
-2 \sigma^2 \pi V_0 \int  \int \phi_0\dpsi i \kv   e^{i \kv \rv} e^{- \sigma^2 k^2}  \frac{\dk}{(2\pi)^2} d \rv = -2 \sigma^2 \pi V_0 \phi_0 \int  \int \int e^{i \rv \kv'} \dpsi_{k'}   i \kv   e^{i \kv \rv} e^{- \sigma^2 k^2}  \frac{\dk}{(2\pi)^4}  \dk' \dr = \\ =-2 \sigma^2 \pi V_0 \phi_0 \int  \int \int e^{  i(\kv' + \kv) \rv} \dpsi_{k'}   i \kv    e^{- \sigma^2 k^2}  \frac{\dk}{(2\pi)^4} \dk' \dr = -2 \sigma^2 \pi V_0 \phi_0 \int   \int \delta(\kv' + \kv) \dpsi_{k'}   i \kv    e^{- \sigma^2 k^2} \frac{\dk}{(2\pi)^4} \dk'  =\\= -2 \sigma^2 \pi V_0 \phi_0   \int  \dpsi_{- k}   i \kv    e^{- \sigma^2 k^2}  \frac{\dk}{(2\pi)^2} =  2 \sigma^4 \pi^2 V_0^2 n   \int  e^{- \sigma^2 k^2} \frac{S(i \Gamma_d - \kv \vv + \omega(\kv) ) + i 2 \mu {\mathcal Im} (\mathcal F (P)) }{(\Gamma_d + i \kv \vv)^2 + 2 \mu  \omega (\kv) + \omega^2}  i \kv \frac{\dk}{(2\pi)^2}.
\end{multline}
Further we calculate the $2$d drag force for $P = 0$ by considering polar coordinates.
\begin{equation}\label{integral1x}
\frac{ F_v}{c} = \int    \frac{ i f_{2 {\rm d} } \omega \kv \vv  e^{- \sigma^2 k^2}  }{(\Gamma_d -  i\kv \vv )^2 + 2 \mu  \omega + \omega^2}   \frac{\dk}{(2\pi)^2} =\int    \frac{ i f_{2 {\rm d} } \omega k v \cos \theta  e^{- \sigma^2 k^2}  }{( \Gamma_d - i k v \cos \theta)^2 + 2 \mu  \omega + \omega^2}  \frac{k}{\sqrt{1 - \cos^2 \theta}} \frac{dk d \cos \theta}{(2\pi)^2} .
\end{equation}
Next we assume a parabolic dispersion and use internal symmetries to simplify the integral, i.e.
\begin{multline}
\int^\infty_0 \int^1_{-1}    \frac{ i f_{2 {\rm d} } k^2 k v \cos \theta  e^{- \sigma^2 k^2}  }{( - i k v \cos \theta + \Gamma_d)^2 + 2 \mu  k^2 + k^4}  \frac{k}{\sqrt{1 - \cos^2 \theta}} \frac{dk d \cos \theta}{(2\pi)^2}  =  \\ =  \int^\infty_0 \int^1_{-1}    \frac{ i f_{2 {\rm d} } k^4 v \cos \theta  e^{- \sigma^2 k^2}  }{-(k v \cos \theta)^2 - i 2 k v \cos \theta\Gamma_d + \Gamma_d^2 + 2 \mu  k^2 + k^4}  \frac{1}{\sqrt{1 - \cos^2 \theta}} \frac{dk d \cos \theta}{(2\pi)^2}  =  \\ =  \int^\infty_0 \int^1_{-1}    \frac{ i f_{2 {\rm d} } k^4 v \cos \theta  e^{- \sigma^2 k^2}  }{( 2k v \cos \theta\Gamma_d)^2 + (-(k v \cos \theta)^2  + \Gamma_d^2 + 2 \mu  k^2 + k^4)^2}  \frac{1}{\sqrt{1 - \cos^2 \theta}} \cdot \left(  i 2 \Gamma_d k v \cos \theta \right) \frac{dk d \cos \theta}{(2\pi)^2}  =  \\ =  - \int^\infty_0 \int^1_{-1}    \frac{ 2 \Gamma_d f_{2 {\rm d} } k^5 v^2 \cos \theta^2  e^{- \sigma^2 k^2}  }{(2 k v \cos \theta\Gamma_d)^2 + (-(k v \cos \theta)^2  + \Gamma_d^2 + 2 \mu  k^2 + k^4)^2}  \frac{1}{\sqrt{1 - \cos^2 \theta}}  \frac{dk d \cos \theta}{(2\pi)^2}.
\end{multline}
Finally we expand the integrand in $v$ to the quadratic order. The result is given by
\beq
\frac{2 e^{-2 \sigma^2 k^2} k^5 x^2 v^2}{(\Gamma_d^2 + k^4 + 2 k^2 \mu)^2 \sqrt{1 - x^2}} + \mathcal O(v^3),
\eeq
where we use the abbreviation $\cos \theta \equiv x$. Integration of $x^2/ \sqrt{1 - x^2}$ gives $\pi/2$ and for $k$ we substitute $k^2 = \rho$ and so obtain 
\beq\label{step}
 - 2v^2 \Gamma_d \int^\infty_0 \int^1_{-1}   \frac{ e^{-2 \sigma^2 \rho} \rho^2 x^2 }{2(\Gamma_d^2 + \rho^2 + 2 \rho \mu)^2 \sqrt{1 - x^2}}   \frac{d \rho d x}{(2\pi)^2} =  - \frac{\pi v^2 \Gamma_d}{2} \int^\infty_0  \frac{ e^{-2 \sigma^2 \rho} \rho^2 }{(\Gamma_d^2 + \rho^2 + 2 \rho \mu)^2}   \frac{d \rho }{(2\pi)^2},
 \eeq
 as presented in the main text.

\subsection{Further example for drag force}

Now we rescale the above expression by $\rho' =  \rho \sigma^2$ for the sake of clarity,
 \begin{multline}\label{nextstep}
\text{\eqref{step}} \to - \frac{\pi v^2 \Gamma_d}{2 \sigma^6} \int^\infty_0  \frac{ e^{-2 \rho'} \rho'^2 }{(\Gamma_d^2 + \rho'^2/\sigma^4 + 2 \rho'/\sigma^2 \mu)^2}   \frac{d \rho' }{(2\pi)^2 } =  - \frac{\pi v^2 \Gamma_d}{2 } \int^\infty_0  \frac{ e^{-2 \rho'} \rho'^2 }{(\sigma^6 \Gamma_d^2 + \sigma^2 \rho'^2 + 2 \sigma^4 \rho' \mu)^2}   \frac{d \rho' }{(2\pi)^2 } = \\ = - \frac{\pi v^2 \Gamma_d}{2 \sigma^2} \int^\infty_0  \frac{ e^{-2 \rho'} \rho'^2 }{(\sigma^4 \Gamma_d^2 + (\rho' + \sigma^2 \mu)^2 - \sigma^4 \mu^2)^2}   \frac{d \rho' }{(2\pi)^2 }.
 \end{multline}
 Finally we write $\rho = \rho' + \sigma^2 \mu$
 \begin{equation}
\text{\eqref{nextstep}} \to - \frac{\pi v^2 \Gamma_d}{2 \sigma^2 (2\pi)^2} e^{ \sigma^2 \mu}  \int^\infty_{\sigma^2 \mu}  \frac{ e^{-2 \rho} (\rho + \sigma^2 \mu)^2 }{( \rho^2  + c_2)^2}d \rho ,
 \end{equation} 
 and define the constant $c_2 = \sigma^4 \Gamma_d^2 - \sigma^4 \mu^2$ and note that the integral is positive. So when $\sigma^2 \mu \simeq 0$ the integral can be solved directly, i.e.
\begin{multline}
\int^\infty_{\sigma^2 \mu \simeq 0} \frac{ e^{-2 \rho} (\rho + \sigma^2 \mu)^2 }{( \rho^2  + c_2)^2}d \rho \simeq \\ \Bigg(\frac{1}{4 c_2 \sqrt{\pi}}(2 \mu^2 {\mathcal MG}[\{\{0\},\{0\}\},\{\{-1/2,0,1\},\{0\}\}, c_2,1] + 2 \sqrt{c_2 \pi} {\mathrm CosIntegral}[2 \sqrt{c_2}] (2 \sqrt{c} \cos[2 \sqrt{c_2}] + + (1 + 4 \mu) \sin{2 \sqrt{c_2}}) +\\  \sqrt{\pi} (\sqrt{c_2} (1 + 4 \mu) \pi \cos{2 \sqrt{c_2}} -  2 (2 \mu + c \pi \sin{2 \sqrt{c_2}} + 2 \sqrt{c_2} (-(1 + 4 \mu) \cos[2 \sqrt{c}] +  2 \sqrt{c_2} \sin{2 \sqrt{c_2}}) {\mathrm SinIntegral}[2 \sqrt{c_2}]) \Bigg)
\end{multline}
with
\begin{multline}
{\mathcal MG}[\{\{a_1, \ldots, a_n\},\{a_{n+1}, \ldots, a_p\}\},\{\{b_1,\ldots,b_m\},\{b_{m+1}, \ldots, b_q\}\}, z,r]  = \\ = \frac{r}{2 \pi i} \int \frac{\Gamma(1-a_1-rs) \ldots \Gamma(1-a_n-rs) \Gamma(b_1+rs) \ldots \Gamma(b_m+rs)}{\Gamma(a_{n+1}+rs) \ldots \Gamma(a_{p}+rs) \Gamma(1-b_{m+1}-rs)\ldots \Gamma(1-b_{q}-rs)} z^{-s} ds
\end{multline}
and $r=1$ and $\Gamma (z) = \int^\infty_0 t^{z-1} e^{-t} dt$.

\subsection{Inhomogeneous dynamics}

Next we turn to the additional pumping terms and we assume $P = \sqrt{n} P_0 \delta (\rv - \vv t)$, while considering again a Gaussian impurity. So by using Eq. $8$ from the main text we obtain the drag force up to the  quadratic order
\begin{equation}
\F =  - \int   |\phi_0 + \dpsi|^2   \nabla V  \dr = - \int   (|\phi_0 |^2  + 2{\mathcal Re}(\phi_0  \dpsi))    \nabla V  \dr + {\mathcal O} (\dpsi^2) = - \int   2{\mathcal Re}(\phi_0  \dpsi)   \nabla V  \dr + {\mathcal O} (\dpsi^2),
\end{equation}
where we used the following identity $ |\phi_0 + \dpsi|^2 = ({\mathcal Re}(\phi_0 + \dpsi))^2 + ({\mathcal Im}(\phi_0 + \dpsi))^2 = |\phi_0|^2 +  2{\mathcal Re}(\phi_0^* \dpsi) + {\mathcal O} (\dpsi^2) $,  again set $\phi_0  \in \mathbb R$ and employed integration between symmetric limits. Using integration by parts, while confining the consideration to $2$d we obtain
\begin{equation}
\F =  2\phi_0  \int   \nabla  {\mathcal Re}( \dpsi) V  \dr + {\mathcal O} (\dpsi^2) =2 \phi_0 {\mathcal Re}\left( \int     \int   \dpsi_{k}   i \kv   e^{i \kv \rv}  \frac{\dk}{(2\pi)^2}V  \dr \right)  + {\mathcal O} (\dpsi^2).
\end{equation}
Then we use common integral algebra to get
\begin{multline}
\F = 2 \phi_0 {\mathcal Re}\left(  \int     \int   \dpsi_{k}   i \kv   e^{i \kv \rv} \frac{\dk}{(2\pi)^2} \left(2 \sigma^2 \pi V_0 \int   e^{i \kv' \rv} e^{- \sigma^2 k'^2}  \frac{\dk'}{(2\pi)} \right)  \dr  \right)  + {\mathcal O} (\dpsi^2)  \simeq  \\ \simeq 2 \cdot  2 \sigma^4 \pi^2 V_0^2 n  \int  e^{- \sigma^2 k^2} {\mathcal Re}\left(  \frac{S(i \Gamma_d + \kv \vv + \omega(\kv) ) + i 2 \mu {\mathcal Im} (\mathcal F (P)) }{(\Gamma_d - i \kv \vv)^2 + 2 \mu  \omega (\kv) + \omega^2}  i \kv \right)  \frac{\dk}{(2\pi)^2} 
\end{multline}
As in the previous section we obtain the transformed expression when assuming $P = \sqrt{n} P_0 \delta (\rv - \vv t)$ and projecting the force on $\vv$ and using symmetric limits. The result is
\beq
\frac{F_v}{c_{\rm Dirac}}=   \int^\infty_0 \int^1_{-1}  \left(  \frac{ i f_{2 {\rm d} } k^4 v \cos \theta - i P_0 \sqrt{n}  \Gamma_d k^2 v \cos \theta }{( -2k v \cos \theta\Gamma_d)^2 + (-(k v \cos \theta)^2  + \Gamma_d^2 + 2 \mu  k^2 + k^4)^2} \left( - i 2 k v \Gamma_d \cos \theta \right) \right)  \frac{e^{- \sigma^2 k^2}}{\sqrt{1 - \cos^2 \theta}} \frac{dk d \cos \theta}{(2\pi)^2}.
\eeq

Furthermore by applying a Taylor expansion in $v$ on the integrand and by considering a Dirac impurity we obtain the additional contribution due to the inhomogeneous terms
\begin{equation}\label{result2}
\frac{F^{\rm Pump}_v}{c_{\rm Dirac}} =  \frac{ P_0 \Gamma^2_d v^2 \pi}{2} \int^\infty_0  \frac{   \rho}{(\Gamma_d^2 + \rho^2 + 2 \rho \mu)^2}   \frac{d \rho }{(2\pi)^2} + {\mathcal O}(v^4) \simeq   \frac{ P_0 \sqrt{n}  \Gamma^2_d v^2 \pi}{2} \frac{1}{4} \left(\frac{2}{d} -\left(\frac{1}{d} \right)^{3/2} \mu \pi + \frac{2 \mu \tan^{-1} (\mu/\sqrt{d})}{d^{3/2}}  \right),
\end{equation}
given $d \equiv \Gamma_d^2  - \mu^2 \neq 0$.

\end{widetext}


\begin{thebibliography}{10}

\bibitem{pita} G. E. Astrakharchik, L.P. Pitaevskii, Phys. Rev. A {\bf 70}, 013608 (2004).

 \bibitem{Pinsk} F. Pinsker, arXiv:1610.04125 (2016).
 
\bibitem{Win} T. Winiecki, J. F. McCann, C. S. Adams, Phys. Rev. Lett. {\bf 82} 26, (1999).
 
 \bibitem{2D1} R. Desbuquois, L. Chomaz, T. Yefsah, J. Léonard, J. Beugnon,
C. Weitenberg, J. Dalibard, Nat. Phys. {\bf 8}, 645 (2012).


\bibitem{gr} M. W. Zwierlein \emph{et al.},  Nature {\bf 442}, 54-58 (2006).

 
 \bibitem{Bat} G. K. Batchelor, {\it An Introduction to Fluid Dynamics}, Cambridge University Press, Cambridge (1967).
 
   \bibitem{puta} L. Pitaevskii and S. Stringari, {\it Bose-Einstein condensation and superfluidity}, Clarendon Press, Oxford (2016).
 
 
\bibitem{Land} L. Landau, Phys. Rev. {\bf 60}, 356 (1941).
 

\bibitem{teo} L. P. Pitaevskii, Journal of Low Temperature Physics, {\bf 87}, Issue 3, pp 127-135 (1992).


 \bibitem{blui}  A. Y. Cherny, J.-S. Caux, J. Brand, Front. Phys., {\bf 7} (1): 54-71 (2012).




\bibitem{uno} P. Kapitza, Nature {\bf 141}, 74,  (1938).

\bibitem{due} J. F. Allen, A. D. Misener, Nature {\bf 141}, 75, (1938).

\bibitem{tre} F. London, Nature {\bf 141}, 643,  (1938).

\bibitem{quattro} A. Einstein, Sitzungsber. Preuss. Akad. Wiss., 3 (1925).

\bibitem{Lon} L. Tisza, Nature {\bf 141}, 913 (1938).


\bibitem{evid} C. Raman \emph{et al.},  Phys. Rev. Lett. {\bf{83}} (13), pp. 2502-2505 (1999).

\bibitem{ketter} R. Onofrio \emph{et al.}, Phys. Rev. Lett. {\bf 85}, 2228-2231 (2000).


\bibitem{heat} C. Raman \emph{et al.}, Journal of Low Temperature Physics, {\bf 122}, pp. 99 (2001).

\bibitem{draggi} D. C. Roberts, Contemporary Physics, {\bf 50}, No. 3, 453-461 (2009).

\bibitem{do1} S. Morgan, Phys. Rev. A {\bf 69}, 023609, (2004).
\bibitem{do2} D. C. Roberts, Phys. Rev. A. {\bf 74}, 013613, (2006).
\bibitem{do3} A. Sykes, M.J. Davis,  D.C. Roberts, Phys. Rev. Lett. {\bf 103} (8), 085302 (2009),
http://arXiv.0904.0995.




\bibitem{frisch} T. Frisch, Y. Pomeau, S. Rica, Phys. Rev. Lett. {\bf{69}}, 1644-1647 (1992).

\bibitem{pom} C. Josserand, Y. Pomeau, S. Rica, Physica D {\bf 134}, 111-125 (1999).

\bibitem{win} T. Winiecki, J. F. McCann,  C. S. Adams, Phys. Rev. Lett. {\bf 82}, 5186-5189 (1999).


\bibitem{Lych1} O. Lychkovskiy, Phys. Rev. A {\bf 91}, 040101(R) (2015).

\bibitem{Lych2} O. Lychkovskiy, Phys. Rev. A {\bf 89}, 033619 (2014).


\bibitem{diss} F. Pinsker, N. G. Berloff,
Phys. Rev. A {\bf 89} (5), 11 (2014).

\bibitem{kasp} J. Kasprzak \emph{et al.}, Nature {\bf 443}, 409-414, doi:10.1038/nature05131 (2006).

\bibitem{exi} A. Amo  \emph{et al.}, Nature Physics {\bf 5}, 805 - 810 (2009).

\bibitem{WouCa} M. Wouters, I. Carusotto, Phys. Rev. Lett. {\bf 105}, 020602 (2010).

\bibitem{Tim}  T. Byrnes, N. Y. Kim, Y. Yamamoto, Nature Physics {\bf 10}, 803-813 (2014).

\bibitem{Sky} E. Cancellieri  \emph{et al.},
Phys. Rev. B {\bf 82}, 224512 (2010).


\bibitem{Hugo} F. Pinsker, H. Flayac, arXiv:1310.7500, Phys. Rev. Lett. {\bf 112} (14), 140405 (2014).

\bibitem{da} V. Goblot \emph{et al.}, arXiv:1607.03711 (2016).

\bibitem{vor} D. Sanvitto \emph{et al.}, Nature Physics {\bf 6}, 527-533 (2010).

\bibitem{vort} K. G. Lagoudakis \emph{et al.}, Nature Physics {\bf 4}, 706-710 (2008).

\bibitem{Baer} J. Keeling, N. G. Berloff, Phys. Rev. Lett. {\bf 100}, 250401 (2008).

\bibitem{Pinsker11} F. Pinsker, Annals of Physics, {\bf 362} pp. 726-738, doi:10.1016/j.aop.2015.09.008 (2015).

\bibitem{F} A.L. Fetter, Phy. Rev. A \textbf{64}, 063608 (2001). 

\bibitem{F1} M. Correggi  \emph{et al.}, Phys. Rev. A \textbf{84},  053614 (2011).

\bibitem{F2} M. Correggi  \emph{et al.}, The European Physical Journal Special Topics 217 (1), 183-188 (2013).


\bibitem{F4} M. Correggi  \emph{et al.}, Journal of Mathematical Physics 53 (9), 095203 (2012).

\bibitem{F5} M. Correggi  \emph{et al.},  J. Stat. Phys. {\bf 143}, 261--305 (2011).

\bibitem{F6} A.L. Fetter, Rev. Mod. Phys. \textbf{81}, 647--691 (2009).

\bibitem{F7} F. Pinsker, {\it Excitations in superfluids of atoms and polaritons}, Department of Applied Mathematics and Theoretical Physics, University of Cambridge (2014).




\bibitem{Light} I. Carusotto, C. Ciuti, Rev. Mod. Phys. {\bf 85}, 299 (2013).

\bibitem{Wout} M. Wouters, I. Carusotto, Phys. Rev. Lett. {\bf 99}, 140402 (2007).

\bibitem{AtomLaser} B. Kneer \emph{et al.}, 
 Phys. Rev. A {\bf 58}, 4841 (1998).


\bibitem{Pita1} L. P. Pitaevskii,  Zh. Eksp. Teor. Fiz. \textbf{40}, 646 (1961); Sov. Phys. JETP \textbf{13}, 451 (1961).

\bibitem{Gross} E. P. Gross, Nuovo Cimento \textbf{20} (1961) 451; J. Math. Phys. \textbf{4} 195 (1963). 

\bibitem{kaka} A. Amo et. al., Journ. of Phys.: Conference Series {\bf 210} 012060 (2010), doi:10.1088/1742-6596/210/1/01206.

\bibitem{leg} A. J. Leggett, 
Rev. Mod. Phys. \textbf{73},  307 (2001).

\bibitem{Inge} E.H. Lieb \emph{et al.}, 
\textit{The Mathematics of the Bose Gas and its Condensation.} Oberwolfach Seminars, {\bf 34}, Birkh\"auser, Basel, $184$pp.  (2001).

\bibitem{alex} A. Dreismann \emph{et al.}, {\bf 111}, 24,  8770-8775 (2014), doi: 10.1073/pnas.1401988111

 \bibitem{PRX} H. Ohadi \emph{et al.}, Phys. Rev. X {\bf 5}, 031002 (2015).
 
  \bibitem{PRL} H. Ohadi \emph{et al.}, Phys. Rev. Lett. {\bf 116}, 106403 (2016).

\bibitem{mk2} F. Pinsker, X. Ruan, T. Alexander, to appear in Scientific Reports
arXiv preprint arXiv:1606.02130 (2016).


\bibitem{Pavloff} N. Pavloff, Phys. Rev. A {\bf 66}, 013610 (2002).







\bibitem{glad} Y. G. Gladush, L.A. Smirnov, A.M. Kamchatnov, J. Phys. B: At. Mol. Opt. Phys. {\bf{41}} 165301 (6pp) (2008).


\bibitem{Mar} A. C. Berceanu, E. Cancellieri, F. M. Marchetti, J. Phys.: Condensed Matter, {\bf 24}, 235802 (2012).


\bibitem{Yukalov} V. I. Yukalov, Laser Physics, {\bf 26}, 062001 (2016).

\bibitem{sok} V. S.  Vladimirov, {\it Equations of mathematical physics}, Marcel Dekker (1971). 



\bibitem{Parameters} Spacial dimensions are assumed to be $2$d. For our numerical integration we use the parameters $n=1$, $\sigma=1$, $V_0=1$, $\Gamma_d = 0.05$, $\mu = 0.04$, then via discretisation of the wave function a global adaptive numerical method yields our simulation. 


\end{thebibliography}
\end{document}